\documentclass[superscriptaddress,amsmath,amssymb,
aps, prstab, twocolumn,groupedaddress]{revtex4-2}
\usepackage{amsmath}
\usepackage{natbib}
\usepackage{float}
\usepackage{amssymb}
\usepackage{dcolumn}
\usepackage{graphicx}
\usepackage{longtable}
\usepackage{bm}
\usepackage{color}
\usepackage{epsfig}
\usepackage{enumerate}

\begin{document}

\title{Large anomalous Hall effect in single crystals of the kagome Weyl ferromagnet Fe$_3$Sn}

\author{Bishnu P. Belbase,$^{1,7,\dag}$ Linda Ye,$^{2,\dag}$ Bishnu Karki,$^{1}$ Jorge I. Facio,$^{3,4}$ Jhih-Shih You,$^{5}$ Joseph G. Checkelsky,$^{2}$ Jeroen van den Brink,$^{6}$ and Madhav Prasad Ghimire,$^{1,6}$}
\email{madhav.ghimire@cdp.tu.edu.np}

\affiliation{$^1$Central Department of Physics, Tribhuvan University, Kirtipur 44613, Kathmandu, Nepal\\
$^2$Department of Physics, Massachusetts Institute of Technology, Cambridge, Massachusetts 02139, USA\\
$^3$ Centro At\'omico Bariloche and Instituto Balseiro, CNEA, 8400 Bariloche, Argentina\\
$^4$ Instituto de Nanociencia y Nanotecnolog\'ia CNEA, 8400 Bariloche, Argentina\\
$^5$Department of Physics, National Taiwan Normal University, Taipei 11677, Taiwan\\
$^6$Leibniz IFW Dresden, Helmholtzstr. 20, Dresden 01069, Germany\\
$^7$Department of Physics and Astronomy, Purdue University, West Lafayette, Indiana 47907, USA}
 
\thanks{These authors contributed equally to this work.}



 




 
\begin{abstract}
The material class of kagome metals has rapidly grown and has been established as a field to explore the interplay between electronic topology and magnetism.
In this work, we report a combined theoretical and experimental study of the anomalous Hall effect of the ferromagnetic kagome metal Fe$_3$Sn.
The compound orders magnetically at 725\,K and presents an easy-plane anisotropy.
Hall measurements in single crystals below room temperature yield an anomalous Hall conductivity $\sigma_{xy}\sim500\,(\Omega\textrm{cm})^{-1}$, which is found to depend weakly on temperature.
This value is in good agreement with the band-intrinsic contribution obtained by density-functional calculations.
Our calculations also yield the correct magnetic anisotropy energy and predict the existence of Weyl nodes near the Fermi energy.
\end{abstract}

\maketitle

\section{Introduction}

In recent years, a number of materials with kagome layers
as structural building block has been investigated. Their electronic structures have been found to host unconventional fermions, such as the Weyl fermions in Mn$_3$(Ge,Si) \cite{nayak2016large,kuroda2017evidence} and Co$_3$Sn$_2$S$_2$ \cite{liu2019magnetic,liu2018giant}, quasi-2D Dirac fermions in Fe$_3$Sn$_2$ \cite{ye2019haas},  and quasi-2D Dirac fermions together with flat bands in FeSn \cite{kang2020dirac}. 

The variety of kagome metals is being intensively studied including compounds based on different transition metals, e.g, FeSn and CoSn \cite{kang2020topological,liu2020orbital,yin2020fermion}, but also different composition variants for a given transition metal, e.g. FeSn and Fe$_3$Sn$_2$.
The two-dimensional limit has been also addressed, e.g., in thin films of Mn$_3$Sn \cite{matsuda2020room}, FeSn \cite{inoue2019molecular} or Co$_3$Sn$_2$S$_2$ \cite{tanaka2020topological} and in monolayers of Cu$_2$Ge, Fe$_2$Ge, and Fe$_2$Sn \cite{liu2019two}. The latter are iso-structural monolayers combining the honeycomb lattice structure of Cu or Fe with the  triangular lattice of Ge or Sn. Interesting electronic properties have been reported, such as  
massive Weyl points in the FM state of Fe$_2$Ge and massless Weyl points in the FM state of Fe$_2$Sn \cite{liu2019two}.

Large efforts have been made to correlate the geometrical aspects of the electronic structure with the electronic response to external fields. In particular, large anomalous Hall and Nernst conductivities have been reported in several kagome metals, both in cases in which the ground state is ferromagnetic \cite{wang2018large,liu2018giant,PhysRevMaterials.4.024202,PhysRevX.9.041061,geishendorf2019signatures} or antiferromagnetic \cite{nakatsuji2015large,nayak2016large,kuroda2017evidence,matsuda2020room}.

In this work, we study the compound Fe$_3$Sn, one natural structural derivative of FeSn and Fe$_3$Sn$_2$ \cite{ye2018massive,kang2020dirac}.
This material has recently attracted attention as a possible permanent magnet due to its large ordering temperature (725\,K) and the possibility of changing its magnetic ground state from easy plane to easy axis through alloying  \cite{Fe3Sn_sales2014ferromagnetism,Fe3Sn_fayyazi2019experimental}.
Recently, large anomalous Nernst data in polycrystals of Fe$_3$Sn have been reported \cite{chen2022large}.
 Here, we report the growth of single crystals along with magnetic properties and measurements of the anomalous Hall resistivity. In addition, we calculate the  electronic structure by means of density-functional theory (DFT) calculations. Our main result is the experimental finding of a large anomalous Hall conductivity (AHC) of about $500\,(\Omega\textrm{cm})^{-1}$, a value which is found to agree well with the band-intrinsic contribution computed from DFT. The measured AHC is weakly temperature dependent below room temperature.
Further, our calculations indicate the existence of Weyl nodes near the Fermi surface and suggest that moderate hole doping could further increase the AHC up to values of $\sim 1250\,(\Omega\textrm{cm})^{-1}$.

This work is organized as follows. Section \ref{methods} describes the experimental and computational methods used in our investigation. Section \ref{exp_results} presents experimental results with focus on the anomalous transport properties. Section \ref{dft_results} contains our theoretical results, which include a characterization of the bulk electronic and magnetic properties, as well as a characterization of the Weyl nodes structure in the ferromagnetic phase.
Last, Section \ref{conclusions} contains our concluding remarks.

\section{Methods}
\label{methods}
\subsection{Experimental Methods}
\begin{figure}[h]
  \includegraphics[width = \columnwidth]{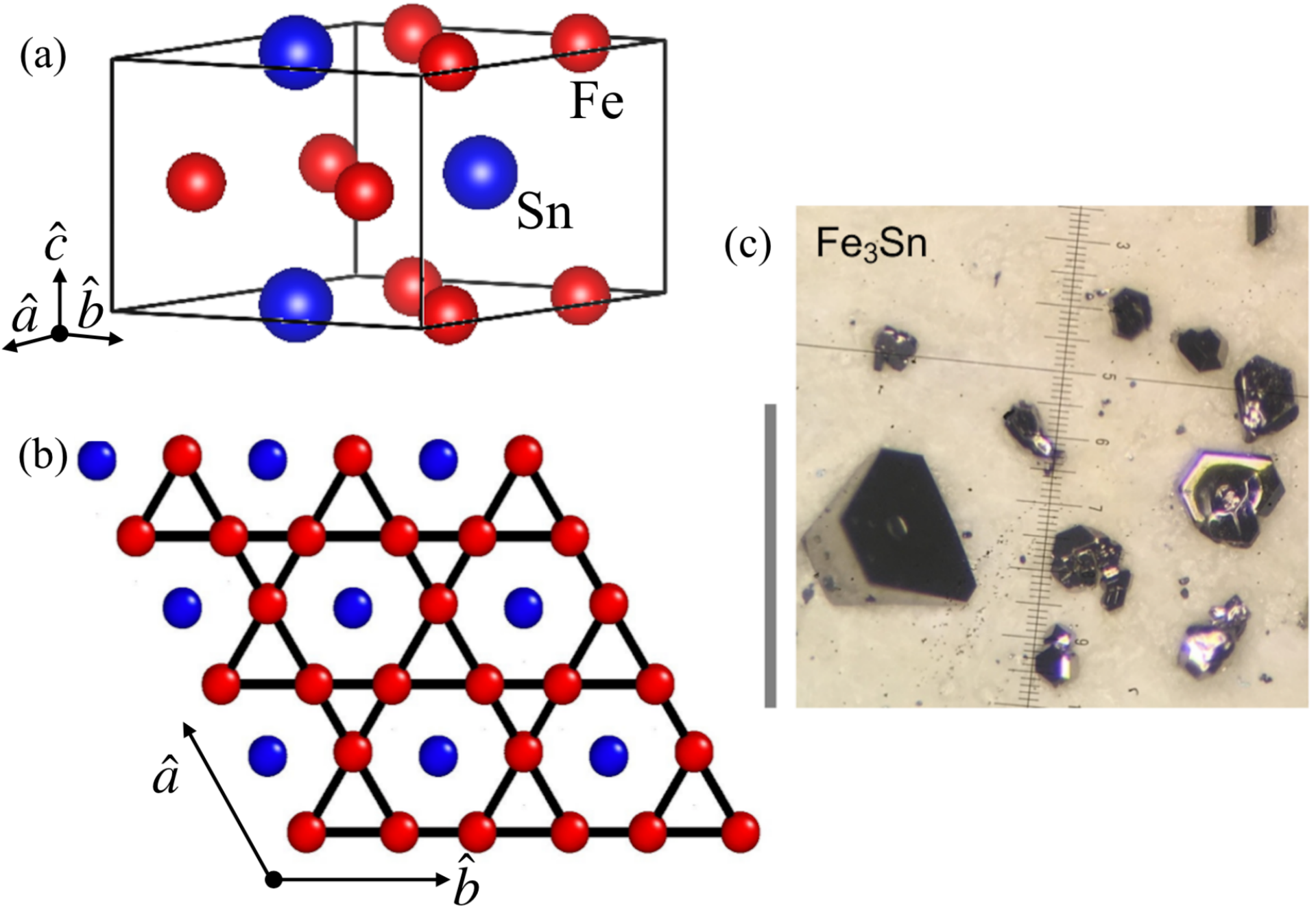}
\caption{(a) Crystal structure of Fe$_3$Sn. (b) Kagome lattice in the $ab-$plane along the $a$-axis. (c) Photo of Fe$_3$Sn single crystals where the hexagonal facets are (001) planes. The vertical scale bar stands for 1 mm.}
\label{fig:structure}
\end{figure}

The bulk crystal structure of Fe$_3$Sn shown in Fig. \ref{fig:structure}(a,b) consists of layers of iron and tin piled in an A-B stacking. The Fe atoms form a kagome lattice with Sn atoms sitting at the center of the hexagon. This structure belongs to space group $P6_3/mmc$ (SG 194) with the experimental lattice constants $a$ = $b$ = 5.487 \AA~and $c$ = 4.31 \AA, respectively \cite{jain2013commentary, buschow1983magneto}.

In view of the narrow stability regime of Fe$_3$Sn in the iron-tin binary phase diagram \cite{FeSn_phasediagram}, we grew single crystals of Fe$_3$Sn via a solid state reaction. Iron powder (Alfa Aesar, 99.998\%) and tin powder (Alfa Aesar, 99.995\%) were mixed in a ratio of $3.5\sim3:1$ and loaded into an evacuated quartz tube. After being heated up to 850 $^{\circ}$C, kept for 120 hours and slowly cooled to 800 $^{\circ}$C over the course of 400 hours, the quartz tube was quenched to cold water and we found crystals of Fe$_3$Sn up to the lateral size of 500 $\mu$m as we show in Fig. \ref{fig:structure}(c). We note that a small number of Fe$_5$Sn$_3$ crystals can also be found in the quartz tubes. We confirmed with X-ray diffraction (XRD) that during the growth process Fe$_3$Sn and Fe$_5$Sn$_3$ tend to crystallize in distinct morphologies: Fe$_3$Sn forms hexagonal prisms (see crystals in Fig. \ref{fig:structure}(c)) and Fe$_5$Sn$_3$ tends to form elongated hexagonal bipyramids. Phase purity of the sample under study is confirmed by powder XRD (see supplemental material, Fig. S1)~\cite{SM}. Electrical transport measurements were performed using the standard five-probe methods in a commercial superconducting magnet. The magnetic field is applied along the $c$-axis and the current within the $ab$-plane.

\subsection{Computational details}

DFT calculations were performed
using the full potential local orbital (FPLO) code version 18.00-52 \cite{koepernik1999full,fplo}.
 The Generalized Gradient Approximation with the parametrization of Perdew, Burke, and Ernzerhof was used \cite{perdew1996generalized}. Relativistic effects were treated in the four component formalism as implemented in FPLO. 
Brillouin zone (BZ) integrals were performed with the tetrahedron method based on a $k$-mesh formed of $24\times24\times24$ subdivisions. The experimental structure was used as detailed above. To ensure an accurate band structure close to the Fermi level, an enhanced basis set as defined in the appendix to Ref. \cite{lejaeghere2016reproducibility} was used. 

Ferromagnetic (FM) configurations were considered with different quantization axes. 
In particular, the magnetic anisotropy energy (MAE) was computed based on calculations with the magnetic moments oriented within the kagome planes ([100] direction) or perpendicular to these layers ([001] direction). 

The study of Weyl nodes, of the surface spectral properties and of the anomalous Hall conductivities are based on a tight-binding model obtained by constructing maximally projected Wannier functions. For this aim, we used the PYFPLO \cite{fplo} module of the FPLO code. The local Wannier basis set is formed of Fe-3$d$, Fe-4$s$, Sn-5$s$ and Sn-5$p$ orbitals.
The search for Weyl nodes was done with a bisection method that detects sources of the Berry curvature field. After a set of candidate nodes is found, the corresponding Chern number associated to each point is computed by integrating the Berry curvature flux through a small sphere centered at the node, as done in Ref. \cite{PhysRevB.93.201101}.

Finally, the band-intrinsic contribution to the anomalous Hall conductivity is calculated using the formula 
\begin{equation}
 \sigma_{\alpha\beta}^A=-\frac{\epsilon_{\alpha\beta\gamma}e^2}{\hbar}\sum_{n}\int_{BZ}\frac{dk}{(2\pi)^3}f_n(k)\Omega_{n,\gamma}(k)
 \end{equation}
where $n$ is a band index, $\Omega_{n,\gamma}$ is the $\gamma$ component of the Berry curvature, $f_n(k)$ is the occupation number of the Bloch state  and $\epsilon_{\alpha\beta\gamma}$ the Levi-Civita tensor \cite{wang2006ab}.
A 300$\times$ 300$\times$ 300 $k$-mesh was used for this calculation.

\section{Experimental results}
\label{exp_results}

  \begin{figure}[!htb]
      \centering
      \includegraphics[width=\columnwidth]{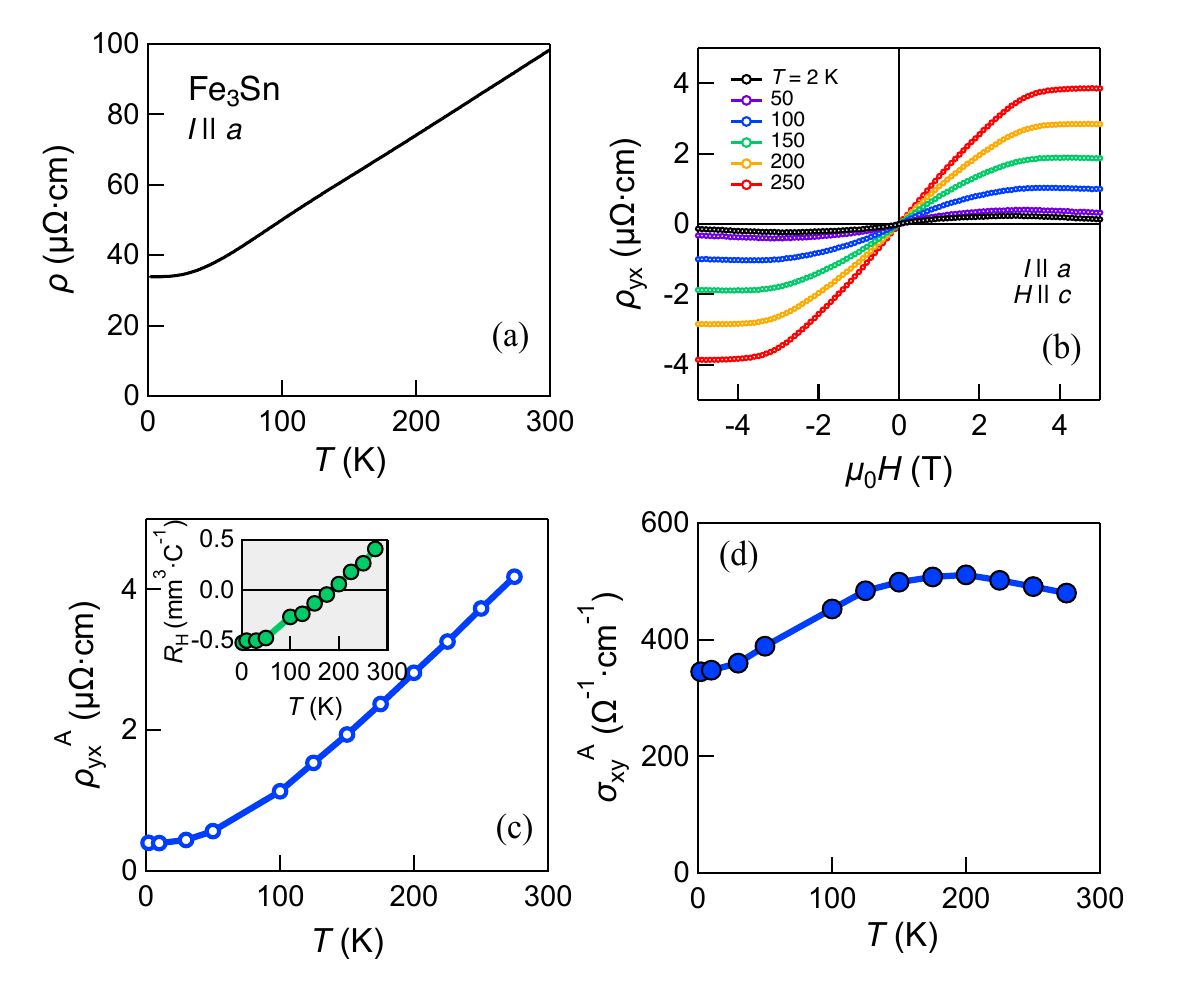}
      \caption{(a) Resistivity $\rho$ of Fe$_3$Sn as a function of $T$ with current $I$ in the kagome lattice plane. (b) Hall resistivity $\rho_{yx}$ as a function of $H$ at selected $T$. Here the magnetic field $H$ is applied along the $c$-axis.  (c) Anomalous Hall resistivity $\rho^A_{yx}$ and ordinary Hall coefficient $R_H$ (inset) vs. $T$ in Fe$_3$Sn extracted from $\rho_{yx}$. (d) $T$-evolution of the anomalous Hall conductivity $\sigma^A_{xy}$ in Fe$_3$Sn.}
      \label{fig:transport}
  \end{figure}

Fe$_3$Sn is known to order ferromagnetically at $T_C=725$ K \cite{Fe3Sn_sales2014ferromagnetism,Fe3Sn_fayyazi2019experimental}. The resistivity $\rho$ of Fe$_3$Sn as a function of temperature $T$ within the kagome lattice plane is shown in Fig. \ref{fig:transport}(a), exhibiting a residual resistivity ratio (RRR) of 2.9. We assume that the diffusive nature of the solid state reaction process facilitates disorder formation throughout the synthesis process, resulting in the high residual resistivity $\sim35$ $\mathrm{\mu\Omega\cdot cm}$ and low RRR in Fe$_3$Sn as compared with it's cousin compounds Fe$_3$Sn$_2$ \cite{ye2018massive} and FeSn \cite{kang2020dirac}.

As expected in prototypical ferromagnetic metals, we observe in the Hall traces of Fe$_3$Sn a prominent anomalous component as shown in Fig. \ref{fig:transport}(b). Saturation is observed at about 3 T external field, which we relate to the complete re-orientation of the magnetization from its easy-plane ground state to the c-axis. The Hall resistivity $\rho_{yx}$ can be decomposed into the ordinary and anomalous contributions via $\rho_{yx}=R_HB_z+R_SM_z$, where $R_H$ is the ordinary Hall coefficient, $B_z$ the magnetic induction in $z$-direction, $R_S$ the anomalous Hall coefficient, and $M_z$ the magnetization in $z$-direction. $R_SM_z$ gives the anomalous Hall resistivity $\rho_{yx}^A$. We infer $\rho_{yx}^A$ and $R_H$ from the high field intercept and slope (above 3 T) of $\rho_{yx}$, respectively. This means, the extracted $\rho_{yx}^A$ data refers to saturated $M_z$. The results are shown in Fig. \ref{fig:transport}(c). $\rho_{yx}^A$ grows monotonically with $T$ while $R_H$ exhibits a sign reversal; $R_H$ at $T= 2$ K corresponds to a carrier density of $1.2\times10^{22}/\mathrm{cm}^3$. The observed $R_H$ suggests the coexistence of both large electron and hole pockets in the system. The anomalous Hall conductivity $\sigma_{xy}^A$ estimated from $\rho_{yx}^A$ via $\sigma_{xy}^A\sim{\rho_{yx}^A}/\rho^2$ is displayed in Fig. \ref{fig:transport}(d). $\sigma_{xy}^A$ is found to depend weakly on $T$ and the maximum is around 500 $\Omega^{-1}\mathrm{cm}^{-1}$. We note that the system is characterized by longitudinal conductivity $\sigma_{xx}\approx1/\rho$ between $1\sim3\times10^4$ $\mathrm{\Omega}^{-1}\mathrm{cm}^{-1}$, and falls rather in the moderately dirty regime, such that the observed $\sigma_{xy}^A$ is expected to reflect an underlying intrinsic Berry curvature contribution  \cite{onoda2008quantum,nagaosa2010anomalous}.  
\section{COMPUTATIONAL RESULTS}
\label{dft_results}

In this section, we first analyze the magnetic properties and electronic structure of Fe$_3$Sn, we report the existence of several Weyl nodes near the Fermi surface and we present calculations of the band-structure-intrinsic contribution to the anomalous Hall conductivity.

 \begin{figure}[!htb]
 	  		\includegraphics[width = \columnwidth]{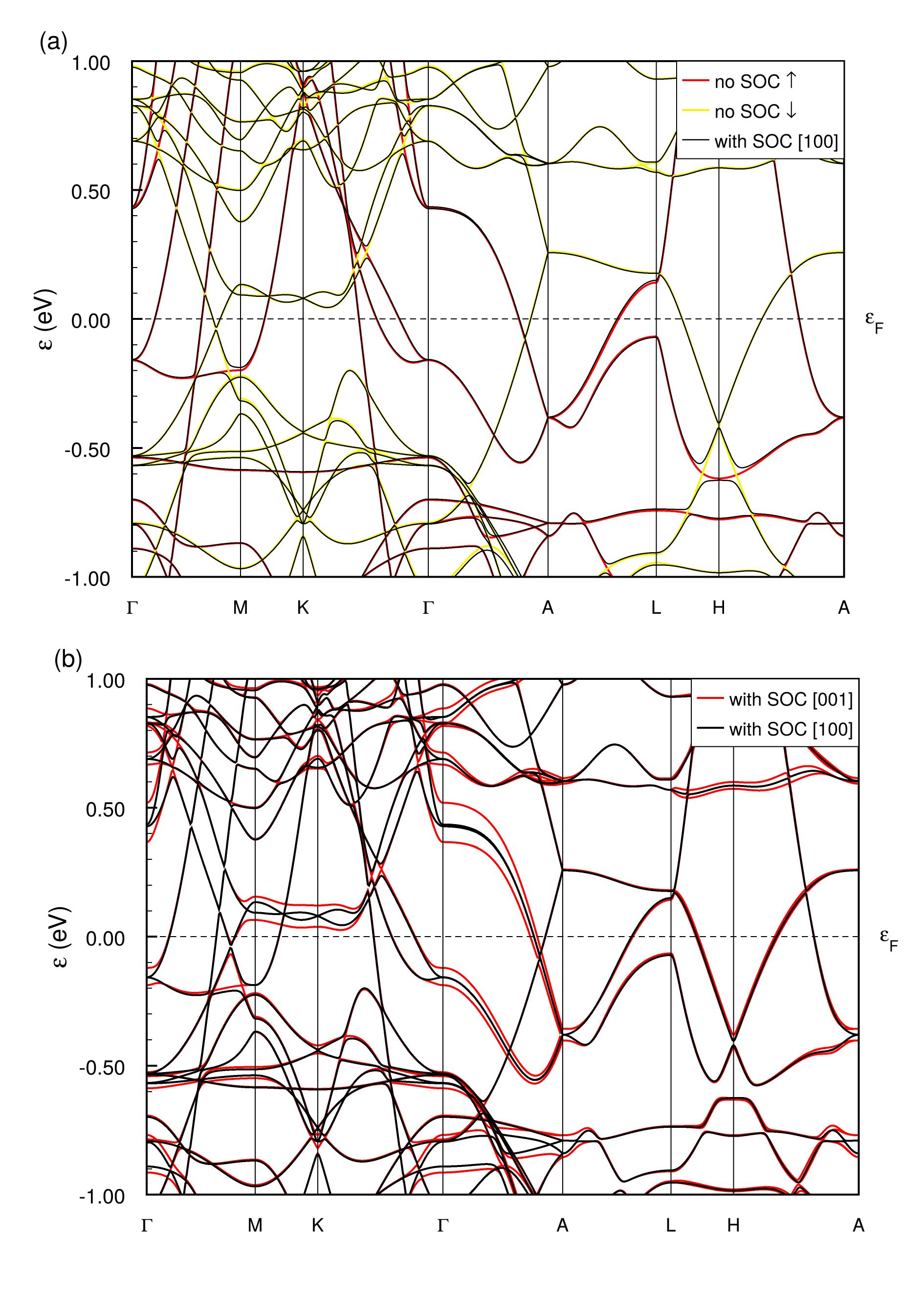}
	 \caption{(a) Band structures in the scalar relativistic and fully-relativistic approximations with the magnetic quantization axis along [100]. (b) Fully relativistic band structures with the quantization axis along [100] and [001].} 
 	\label{bands}
 \end{figure}

  Table \ref{table1} shows the total energies and magnetic moments obtained for different quantization axes. The FM state with quantization axis along the [100] direction is found to be the ground state with a total magnetic moment of 7.22$\mu_B$ per Fe$_3$Sn. 

The magnetic moment is slightly reduced if the magnetization is oriented along the c-axis. The MAE calculated from the energy difference between these two states amounts to $\sim0.64$\,meV per Fe$_3$Sn. Note that we refrain from an analysis of in-plane magnetic anisotropy, which is very small in hexagonal systems and not expected to affect the transport properties considered here. Under the assumption of negligible higher-order terms, this translates into an anisotropy constant $K_1$ = - 1.8\,MJ$/$m$^3$ and an anisotropy field of 3.1 T. The former value agrees with earlier theoretical results of $\lvert K_1\rvert$ = 1.59\,MJ$/$m$^3$ \cite{Fe3Sn_sales2014ferromagnetism} (the authors of this reference use a sign convention opposite to the usual one) and of - 1.5\,MJ$/$m$^3$ \cite{PhysRevB.99.024421}. The calculated anisotropy field agrees with the experimentally observed saturation around 3 T, see Fig. 2 (b), and with the same result of 3 T obtained from earlier magnetization data \cite{Fe3Sn_sales2014ferromagnetism}. We tried the application of so-called orbital polarization corrections (OPC) \cite{nordstrom1992calculation} which, however, yield a two-fold enhanced anisotropy field. Thus, we conclude that Fe$_3$Sn is better described by conventional GGA than by GGA-OPC and use only GGA data for the further investigation.
  
 \begin{table}[!htb]
\centering
 \caption{ Total energy referred to the ground state energy and total magnetic moments (including spin and orbital contributions) for the two
considered magnetic states. The reported data are calculated considering the extended basis set.}
 
\begin{tabular}{|c|c|c|}
\hline
\hline
State & Energy (meV / f.u.) & Total moment($\mu_B$ / f.u.) \\ \hline
 FM [100]& 0 &  7.22 \\  
 FM [001]& 0.64    &  7.20 \\
\hline
\hline
\end{tabular}
\label{table1}
\end{table}
 
\begin{figure}[!htb]
  \includegraphics[width = \columnwidth]{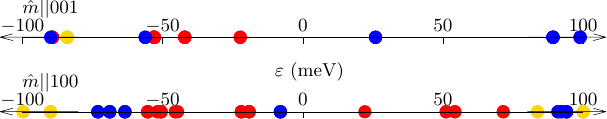}
\caption{Energy of Weyl nodes near the Fermi surface for the magnetization along the [001] and [100] directions. Red, blue and orange dots correspond to Weyl nodes in which the lower band has index $N-1$, $N$ and $N+1$, respectively, with $N$ the number of valence electrons.}
\label{fig:wp}
\end{figure}

 Figure \ref{bands}(a) presents the energy dispersion obtained in the scalar and in the fully-relativistic calculations with [100] orientation of the magnetization. Several bands, which mainly present Fe-$3d$ character, cross the Fermi energy. Bands with relatively flat energy dispersion as a fingerprint of the kagome lattice are present in the $\Gamma$-$M$-$K$ plane around -0.5\,eV as well as in the $A$-$L$-$H$ plane around +0.6\,eV and around -0.8\,eV. Without SOC, apparent low-energy band crossings along the paths $\Gamma$-M and  M-K-$\Gamma$ can be observed, which appear to be robust to the SOC for this direction of the magnetic moments. Fig. \ref{bands}(b) compares the bands with SOC for the FM [001] and FM [100] configurations and shows that these crossings become gapped when the magnetization points along [001].\\ 
A search of Weyl nodes for the FM [100] ground state yields the existence of several Weyl nodes, the closest to Fermi surface are found 8\,meV below the Fermi energy. 
Notice that in SG 194, for this magnetic configuration, there can be up to eight symmetry-related Weyl nodes generated by the inversion symmetry $I$, the two-fold rotations $C_2(y)\times \Theta$, $C_2(2x+y)$, $C_2(z)\times \Theta$ and the reflection symmetries $m(y) \times \Theta$, $m(2x+y)$ and  $m(z)\times \Theta$, where $\Theta$ is the time-reversal operator.\\
When the magnetization is rotated towards the hard axis, a situation experimentally relevant in view of the MAE found above and the observed saturation of $\rho_{yx}$, one can expect the Weyl nodes to move in momentum and energy space as theoretically found in other magnetic compounds \cite{ghimire2019creating,ray2022tunable}. A search of Weyl nodes for the FM [001] indeed leads to sizeable changes in the number and energy of the Weyl nodes near the Fermi energy, as illustrated in Fig. \ref{fig:wp}, which shows the found Weyl nodes ordered according to their energy, relative to the Fermi energy, for both magnetic configurations. We also present the distribution of these Weyl nodes in the Brillouin zone in the supplemental material (see Fig. S2 and S3)~\cite{SM}.
 
\begin{figure}[!htb]
	\includegraphics[width=3.4in, height= 2.5in]{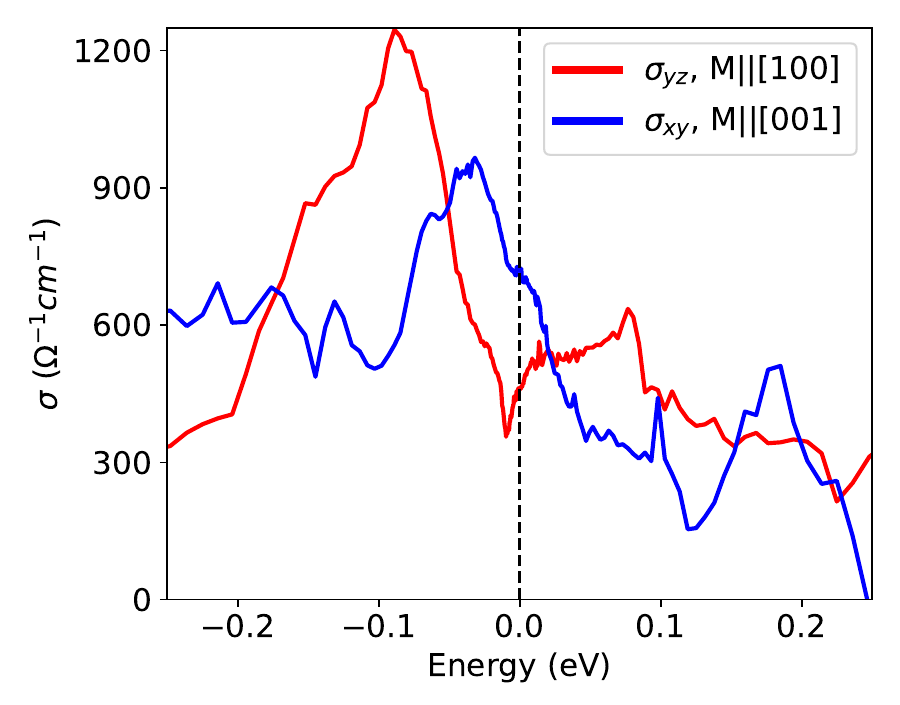}
	\caption{Anomalous Hall conductivity as a function of the chemical potential.}
	\label{fig:ahc}
\end{figure}
Fig. \ref{fig:ahc} presents the band-intrinsic contribution to the anomalous Hall conductivity as a function of the chemical potential. For quantization axis along [001] ([100]) only $\sigma_{xy}$ ($\sigma_{yz}$) is different from zero.
At the charge neutrality point ($\mu=0$), the related values are $\sigma_{xy}^A=702\,(\Omega \textrm{cm})^{-1}$ and $\sigma_{yz}^A=482\,(\Omega \textrm {cm})^{-1}$. The former value is in fairly good agreement with the experimental result. In the supplemental material~\cite{SM}, we analyze the contribution to the AHC from states at different parts of the Brillouin zone and show that the most prominent hot spots acquire a loop structure which reflects the importance of structures other than isolated Weyl nodes for the AHC. The AHC obtained for Fe$_3$Sn is significantly larger than Fe$_3$Sn$_2$ with a value of 200 $(\Omega \textrm {cm})^{-1}$~\cite{ye2018massive}, and is comparable to Fe$_5$Sn$_3$ with a value of 507 $(\Omega \textrm {cm})^{-1}$~\cite{li2020large}.
Interestingly, according to the calculations a moderate amount of hole doping could increase $\sigma_{xy}^A$ and $\sigma_{yz}^A$ up to maximum values of $\sim950\,(\Omega \textrm {cm})^{-1}$ and $\sim1250\,(\Omega \textrm {cm})^{-1}$, respectively. However, this increase is contingent upon maintaining the stability of the ground state property during the doping process~\cite{ozawa2022self,guguchia2021multiple}.  
Recent values of AHC at $E_{\textrm F}$~\cite{shen2022thermodynamical} is found to be larger than our reported value obtained from the enhanced basis calculations, but is well supported with our earlier work~\cite{ghimire2021large}.
 
\section{Conclusions}
\label{conclusions}

We have studied single crystalline kagome metal Fe$_3$Sn by means of transport experiments and on the basis of density-functional calculations. The system is found to be ferromagnetic, with an easy-plane magnetic anisotropy. A moderate magnetic field of about 3\,T is sufficient to reorientate the magnetization along the hexagonal axis. From transport measurements we have inferred the anomalous Hall conductivity to be $\sim500\,(\Omega \textrm {cm})^{-1}$. Our DFT results reproduce the preference of the easy-plane anisotropy and our calculation of the band-intrinsic contribution to the AHC agrees well with the experimental results. Interestingly, our calculations indicate that moderate hole doping could largely increase the AHC. All-together, our experimental and theoretical results establish  Fe$_3$Sn as a kagome Weyl metal with large AHC.

\acknowledgments
We thank Manuel Richter for fruitful discussions, suggestions and for reading the manuscript. We also thank Ulrike Nitzsche for technical assistance.
M.P.G. thanks the Alexander von Humboldt-Stiftung for the financial support through equipment grants and renewed research program. M.P.G. was also supported by a grant from UNESCO-TWAS and the Swedish International Development Cooperation Agency (SIDA). The views expressed herein do not necessarily represent those of UNESCO-TWAS, SIDA or its Board of Governors.  J.I.F. would like to acknowledge the support from the Alexander von Humboldt-Stiftung and ANPCyT grants PICT 2018/01509 and PICT 2019/00371. L.Y. acknowledges support by the Tsinghua Education Foundation and STC Center for Integrated Quantum Materials, NSF grant number DMR-1231319. This work was funded, in part, by the Gordon and Betty Moore Foundation EPiQS Initiative, through Grants GBMF3848 and GBMF9070 to J.G.C. (material synthesis) and NSF grant DMR-2104964 (material analysis). J.-S.Y. is supported by the National Science and Technology Council (Grant No. MOST 110-2112-M-003-008-MY3) and National Center for Theoretical Sciences in Taiwan. J.v.d.B. acknowledges financial support from the German Research Foundation (Deutsche Forschungsgemeinschaft, DFG) via SFB 1143 Project No. A5 and under Germany's Excellence Strategy through the W{\"u}rzburg-Dresden Cluster of Excellence on Complexity and Topology in Quantum Matter ct.qmat (EXC 2147, Project No. 390858490).
\bibliographystyle{apsrev}
\bibliography{references_v1}

\begin{thebibliography}{43}
\expandafter\ifx\csname natexlab\endcsname\relax\def\natexlab#1{#1}\fi
\expandafter\ifx\csname bibnamefont\endcsname\relax
  \def\bibnamefont#1{#1}\fi
\expandafter\ifx\csname bibfnamefont\endcsname\relax
  \def\bibfnamefont#1{#1}\fi
\expandafter\ifx\csname citenamefont\endcsname\relax
  \def\citenamefont#1{#1}\fi
\expandafter\ifx\csname url\endcsname\relax
  \def\url#1{\texttt{#1}}\fi
\expandafter\ifx\csname urlprefix\endcsname\relax\def\urlprefix{URL }\fi
\providecommand{\bibinfo}[2]{#2}
\providecommand{\eprint}[2][]{\url{#2}}

\bibitem[{\citenamefont{Nayak et~al.}(2016)\citenamefont{Nayak, Fischer, Sun,
  Yan, Karel, Komarek, Shekhar, Kumar, Schnelle, K{\"u}bler
  et~al.}}]{nayak2016large}
\bibinfo{author}{\bibfnamefont{A.~K.} \bibnamefont{Nayak}},
  \bibinfo{author}{\bibfnamefont{J.~E.} \bibnamefont{Fischer}},
  \bibinfo{author}{\bibfnamefont{Y.}~\bibnamefont{Sun}},
  \bibinfo{author}{\bibfnamefont{B.}~\bibnamefont{Yan}},
  \bibinfo{author}{\bibfnamefont{J.}~\bibnamefont{Karel}},
  \bibinfo{author}{\bibfnamefont{A.~C.} \bibnamefont{Komarek}},
  \bibinfo{author}{\bibfnamefont{C.}~\bibnamefont{Shekhar}},
  \bibinfo{author}{\bibfnamefont{N.}~\bibnamefont{Kumar}},
  \bibinfo{author}{\bibfnamefont{W.}~\bibnamefont{Schnelle}},
  \bibinfo{author}{\bibfnamefont{J.}~\bibnamefont{K{\"u}bler}},
  \bibnamefont{et~al.}, \bibinfo{journal}{Sci. Adv.}
  \textbf{\bibinfo{volume}{2}}, \bibinfo{pages}{e1501870}
  (\bibinfo{year}{2016}).

\bibitem[{\citenamefont{Kuroda et~al.}(2017)\citenamefont{Kuroda, Tomita,
  Suzuki, Bareille, Nugroho, Goswami, Ochi, Ikhlas, Nakayama, Akebi
  et~al.}}]{kuroda2017evidence}
\bibinfo{author}{\bibfnamefont{K.}~\bibnamefont{Kuroda}},
  \bibinfo{author}{\bibfnamefont{T.}~\bibnamefont{Tomita}},
  \bibinfo{author}{\bibfnamefont{M.-T.} \bibnamefont{Suzuki}},
  \bibinfo{author}{\bibfnamefont{C.}~\bibnamefont{Bareille}},
  \bibinfo{author}{\bibfnamefont{A.}~\bibnamefont{Nugroho}},
  \bibinfo{author}{\bibfnamefont{P.}~\bibnamefont{Goswami}},
  \bibinfo{author}{\bibfnamefont{M.}~\bibnamefont{Ochi}},
  \bibinfo{author}{\bibfnamefont{M.}~\bibnamefont{Ikhlas}},
  \bibinfo{author}{\bibfnamefont{M.}~\bibnamefont{Nakayama}},
  \bibinfo{author}{\bibfnamefont{S.}~\bibnamefont{Akebi}},
  \bibnamefont{et~al.}, \bibinfo{journal}{Nat. Mater.}
  \textbf{\bibinfo{volume}{16}}, \bibinfo{pages}{1090} (\bibinfo{year}{2017}).

\bibitem[{\citenamefont{Liu et~al.}(2019)\citenamefont{Liu, Liang, Liu, Xu, Li,
  Chen, Pei, Shi, Mo, Dudin et~al.}}]{liu2019magnetic}
\bibinfo{author}{\bibfnamefont{D.}~\bibnamefont{Liu}},
  \bibinfo{author}{\bibfnamefont{A.}~\bibnamefont{Liang}},
  \bibinfo{author}{\bibfnamefont{E.}~\bibnamefont{Liu}},
  \bibinfo{author}{\bibfnamefont{Q.}~\bibnamefont{Xu}},
  \bibinfo{author}{\bibfnamefont{Y.}~\bibnamefont{Li}},
  \bibinfo{author}{\bibfnamefont{C.}~\bibnamefont{Chen}},
  \bibinfo{author}{\bibfnamefont{D.}~\bibnamefont{Pei}},
  \bibinfo{author}{\bibfnamefont{W.}~\bibnamefont{Shi}},
  \bibinfo{author}{\bibfnamefont{S.}~\bibnamefont{Mo}},
  \bibinfo{author}{\bibfnamefont{P.}~\bibnamefont{Dudin}},
  \bibnamefont{et~al.}, \bibinfo{journal}{Science}
  \textbf{\bibinfo{volume}{365}}, \bibinfo{pages}{1282} (\bibinfo{year}{2019}).

\bibitem[{\citenamefont{Liu et~al.}(2018)\citenamefont{Liu, Sun, Kumar,
  Muechler, Sun, Jiao, Yang, Liu, Liang, Xu et~al.}}]{liu2018giant}
\bibinfo{author}{\bibfnamefont{E.}~\bibnamefont{Liu}},
  \bibinfo{author}{\bibfnamefont{Y.}~\bibnamefont{Sun}},
  \bibinfo{author}{\bibfnamefont{N.}~\bibnamefont{Kumar}},
  \bibinfo{author}{\bibfnamefont{L.}~\bibnamefont{Muechler}},
  \bibinfo{author}{\bibfnamefont{A.}~\bibnamefont{Sun}},
  \bibinfo{author}{\bibfnamefont{L.}~\bibnamefont{Jiao}},
  \bibinfo{author}{\bibfnamefont{S.-Y.} \bibnamefont{Yang}},
  \bibinfo{author}{\bibfnamefont{D.}~\bibnamefont{Liu}},
  \bibinfo{author}{\bibfnamefont{A.}~\bibnamefont{Liang}},
  \bibinfo{author}{\bibfnamefont{Q.}~\bibnamefont{Xu}}, \bibnamefont{et~al.},
  \bibinfo{journal}{Nat. Phys.} \textbf{\bibinfo{volume}{14}},
  \bibinfo{pages}{1125} (\bibinfo{year}{2018}).

\bibitem[{\citenamefont{Ye et~al.}(2019)\citenamefont{Ye, Chan, McDonald, Graf,
  Kang, Liu, Suzuki, Comin, Fu, and Checkelsky}}]{ye2019haas}
\bibinfo{author}{\bibfnamefont{L.}~\bibnamefont{Ye}},
  \bibinfo{author}{\bibfnamefont{M.~K.} \bibnamefont{Chan}},
  \bibinfo{author}{\bibfnamefont{R.~D.} \bibnamefont{McDonald}},
  \bibinfo{author}{\bibfnamefont{D.}~\bibnamefont{Graf}},
  \bibinfo{author}{\bibfnamefont{M.}~\bibnamefont{Kang}},
  \bibinfo{author}{\bibfnamefont{J.}~\bibnamefont{Liu}},
  \bibinfo{author}{\bibfnamefont{T.}~\bibnamefont{Suzuki}},
  \bibinfo{author}{\bibfnamefont{R.}~\bibnamefont{Comin}},
  \bibinfo{author}{\bibfnamefont{L.}~\bibnamefont{Fu}}, \bibnamefont{and}
  \bibinfo{author}{\bibfnamefont{J.~G.} \bibnamefont{Checkelsky}},
  \bibinfo{journal}{Nat. Commun.} \textbf{\bibinfo{volume}{10}},
  \bibinfo{pages}{1} (\bibinfo{year}{2019}).

\bibitem[{\citenamefont{Kang et~al.}(2020{\natexlab{a}})\citenamefont{Kang, Ye,
  Fang, You, Levitan, Han, Facio, Jozwiak, Bostwick, Rotenberg
  et~al.}}]{kang2020dirac}
\bibinfo{author}{\bibfnamefont{M.}~\bibnamefont{Kang}},
  \bibinfo{author}{\bibfnamefont{L.}~\bibnamefont{Ye}},
  \bibinfo{author}{\bibfnamefont{S.}~\bibnamefont{Fang}},
  \bibinfo{author}{\bibfnamefont{J.-S.} \bibnamefont{You}},
  \bibinfo{author}{\bibfnamefont{A.}~\bibnamefont{Levitan}},
  \bibinfo{author}{\bibfnamefont{M.}~\bibnamefont{Han}},
  \bibinfo{author}{\bibfnamefont{J.~I.} \bibnamefont{Facio}},
  \bibinfo{author}{\bibfnamefont{C.}~\bibnamefont{Jozwiak}},
  \bibinfo{author}{\bibfnamefont{A.}~\bibnamefont{Bostwick}},
  \bibinfo{author}{\bibfnamefont{E.}~\bibnamefont{Rotenberg}},
  \bibnamefont{et~al.}, \bibinfo{journal}{Nat. Mater.}
  \textbf{\bibinfo{volume}{19}}, \bibinfo{pages}{163}
  (\bibinfo{year}{2020}{\natexlab{a}}).

\bibitem[{\citenamefont{Kang et~al.}(2020{\natexlab{b}})\citenamefont{Kang,
  Fang, Ye, Po, Denlinger, Jozwiak, Bostwick, Rotenberg, Kaxiras, Checkelsky
  et~al.}}]{kang2020topological}
\bibinfo{author}{\bibfnamefont{M.}~\bibnamefont{Kang}},
  \bibinfo{author}{\bibfnamefont{S.}~\bibnamefont{Fang}},
  \bibinfo{author}{\bibfnamefont{L.}~\bibnamefont{Ye}},
  \bibinfo{author}{\bibfnamefont{H.~C.} \bibnamefont{Po}},
  \bibinfo{author}{\bibfnamefont{J.}~\bibnamefont{Denlinger}},
  \bibinfo{author}{\bibfnamefont{C.}~\bibnamefont{Jozwiak}},
  \bibinfo{author}{\bibfnamefont{A.}~\bibnamefont{Bostwick}},
  \bibinfo{author}{\bibfnamefont{E.}~\bibnamefont{Rotenberg}},
  \bibinfo{author}{\bibfnamefont{E.}~\bibnamefont{Kaxiras}},
  \bibinfo{author}{\bibfnamefont{J.~G.} \bibnamefont{Checkelsky}},
  \bibnamefont{et~al.}, \bibinfo{journal}{Nat. Commun.}
  \textbf{\bibinfo{volume}{11}}, \bibinfo{pages}{4004}
  (\bibinfo{year}{2020}{\natexlab{b}}).

\bibitem[{\citenamefont{Liu et~al.}(2020{\natexlab{a}})\citenamefont{Liu, Li,
  Wang, Wang, Wen, Jiang, Lu, Yan, Huang, Shen et~al.}}]{liu2020orbital}
\bibinfo{author}{\bibfnamefont{Z.}~\bibnamefont{Liu}},
  \bibinfo{author}{\bibfnamefont{M.}~\bibnamefont{Li}},
  \bibinfo{author}{\bibfnamefont{Q.}~\bibnamefont{Wang}},
  \bibinfo{author}{\bibfnamefont{G.}~\bibnamefont{Wang}},
  \bibinfo{author}{\bibfnamefont{C.}~\bibnamefont{Wen}},
  \bibinfo{author}{\bibfnamefont{K.}~\bibnamefont{Jiang}},
  \bibinfo{author}{\bibfnamefont{X.}~\bibnamefont{Lu}},
  \bibinfo{author}{\bibfnamefont{S.}~\bibnamefont{Yan}},
  \bibinfo{author}{\bibfnamefont{Y.}~\bibnamefont{Huang}},
  \bibinfo{author}{\bibfnamefont{D.}~\bibnamefont{Shen}}, \bibnamefont{et~al.},
  \bibinfo{journal}{Nat. Commun.} \textbf{\bibinfo{volume}{11}},
  \bibinfo{pages}{4002} (\bibinfo{year}{2020}{\natexlab{a}}).

\bibitem[{\citenamefont{Yin et~al.}(2020)\citenamefont{Yin, Shumiya, Mardanya,
  Wang, Zhang, Tien, Multer, Jiang, Cheng, Yao et~al.}}]{yin2020fermion}
\bibinfo{author}{\bibfnamefont{J.-X.} \bibnamefont{Yin}},
  \bibinfo{author}{\bibfnamefont{N.}~\bibnamefont{Shumiya}},
  \bibinfo{author}{\bibfnamefont{S.}~\bibnamefont{Mardanya}},
  \bibinfo{author}{\bibfnamefont{Q.}~\bibnamefont{Wang}},
  \bibinfo{author}{\bibfnamefont{S.~S.} \bibnamefont{Zhang}},
  \bibinfo{author}{\bibfnamefont{H.-J.} \bibnamefont{Tien}},
  \bibinfo{author}{\bibfnamefont{D.}~\bibnamefont{Multer}},
  \bibinfo{author}{\bibfnamefont{Y.}~\bibnamefont{Jiang}},
  \bibinfo{author}{\bibfnamefont{G.}~\bibnamefont{Cheng}},
  \bibinfo{author}{\bibfnamefont{N.}~\bibnamefont{Yao}}, \bibnamefont{et~al.},
  \bibinfo{journal}{Nat. Commun.} \textbf{\bibinfo{volume}{11}},
  \bibinfo{pages}{4003} (\bibinfo{year}{2020}).

\bibitem[{\citenamefont{Matsuda et~al.}(2020)\citenamefont{Matsuda, Kanda,
  Higo, Armitage, Nakatsuji, and Matsunaga}}]{matsuda2020room}
\bibinfo{author}{\bibfnamefont{T.}~\bibnamefont{Matsuda}},
  \bibinfo{author}{\bibfnamefont{N.}~\bibnamefont{Kanda}},
  \bibinfo{author}{\bibfnamefont{T.}~\bibnamefont{Higo}},
  \bibinfo{author}{\bibfnamefont{N.}~\bibnamefont{Armitage}},
  \bibinfo{author}{\bibfnamefont{S.}~\bibnamefont{Nakatsuji}},
  \bibnamefont{and}
  \bibinfo{author}{\bibfnamefont{R.}~\bibnamefont{Matsunaga}},
  \bibinfo{journal}{Nat. Commun.} \textbf{\bibinfo{volume}{11}},
  \bibinfo{pages}{909} (\bibinfo{year}{2020}).

\bibitem[{\citenamefont{Inoue et~al.}(2019)\citenamefont{Inoue, Han, Ye,
  Suzuki, and Checkelsky}}]{inoue2019molecular}
\bibinfo{author}{\bibfnamefont{H.}~\bibnamefont{Inoue}},
  \bibinfo{author}{\bibfnamefont{M.}~\bibnamefont{Han}},
  \bibinfo{author}{\bibfnamefont{L.}~\bibnamefont{Ye}},
  \bibinfo{author}{\bibfnamefont{T.}~\bibnamefont{Suzuki}}, \bibnamefont{and}
  \bibinfo{author}{\bibfnamefont{J.~G.} \bibnamefont{Checkelsky}},
  \bibinfo{journal}{Appl. Phys. Lett.} \textbf{\bibinfo{volume}{115}},
  \bibinfo{pages}{072403} (\bibinfo{year}{2019}).

\bibitem[{\citenamefont{Tanaka et~al.}(2020)\citenamefont{Tanaka, Fujishiro,
  Mogi, Kaneko, Yokosawa, Kanazawa, Minami, Koretsune, Arita, Tarucha
  et~al.}}]{tanaka2020topological}
\bibinfo{author}{\bibfnamefont{M.}~\bibnamefont{Tanaka}},
  \bibinfo{author}{\bibfnamefont{Y.}~\bibnamefont{Fujishiro}},
  \bibinfo{author}{\bibfnamefont{M.}~\bibnamefont{Mogi}},
  \bibinfo{author}{\bibfnamefont{Y.}~\bibnamefont{Kaneko}},
  \bibinfo{author}{\bibfnamefont{T.}~\bibnamefont{Yokosawa}},
  \bibinfo{author}{\bibfnamefont{N.}~\bibnamefont{Kanazawa}},
  \bibinfo{author}{\bibfnamefont{S.}~\bibnamefont{Minami}},
  \bibinfo{author}{\bibfnamefont{T.}~\bibnamefont{Koretsune}},
  \bibinfo{author}{\bibfnamefont{R.}~\bibnamefont{Arita}},
  \bibinfo{author}{\bibfnamefont{S.}~\bibnamefont{Tarucha}},
  \bibnamefont{et~al.}, \bibinfo{journal}{Nano Lett.}
  \textbf{\bibinfo{volume}{20}}, \bibinfo{pages}{7476} (\bibinfo{year}{2020}).

\bibitem[{\citenamefont{Liu et~al.}(2020{\natexlab{b}})\citenamefont{Liu, Wang,
  Li, Chen, Jia, and Cho}}]{liu2019two}
\bibinfo{author}{\bibfnamefont{L.}~\bibnamefont{Liu}},
  \bibinfo{author}{\bibfnamefont{C.}~\bibnamefont{Wang}},
  \bibinfo{author}{\bibfnamefont{J.}~\bibnamefont{Li}},
  \bibinfo{author}{\bibfnamefont{X.-Q.} \bibnamefont{Chen}},
  \bibinfo{author}{\bibfnamefont{Y.}~\bibnamefont{Jia}}, \bibnamefont{and}
  \bibinfo{author}{\bibfnamefont{J.-H.} \bibnamefont{Cho}},
  \bibinfo{journal}{Phys. Rev. B} \textbf{\bibinfo{volume}{101}},
  \bibinfo{pages}{165403} (\bibinfo{year}{2020}{\natexlab{b}}).

\bibitem[{\citenamefont{Wang et~al.}(2018)\citenamefont{Wang, Xu, Lou, Liu, Li,
  Huang, Shen, Weng, Wang, and Lei}}]{wang2018large}
\bibinfo{author}{\bibfnamefont{Q.}~\bibnamefont{Wang}},
  \bibinfo{author}{\bibfnamefont{Y.}~\bibnamefont{Xu}},
  \bibinfo{author}{\bibfnamefont{R.}~\bibnamefont{Lou}},
  \bibinfo{author}{\bibfnamefont{Z.}~\bibnamefont{Liu}},
  \bibinfo{author}{\bibfnamefont{M.}~\bibnamefont{Li}},
  \bibinfo{author}{\bibfnamefont{Y.}~\bibnamefont{Huang}},
  \bibinfo{author}{\bibfnamefont{D.}~\bibnamefont{Shen}},
  \bibinfo{author}{\bibfnamefont{H.}~\bibnamefont{Weng}},
  \bibinfo{author}{\bibfnamefont{S.}~\bibnamefont{Wang}}, \bibnamefont{and}
  \bibinfo{author}{\bibfnamefont{H.}~\bibnamefont{Lei}}, \bibinfo{journal}{Nat.
  Commun.} \textbf{\bibinfo{volume}{9}}, \bibinfo{pages}{1}
  (\bibinfo{year}{2018}).

\bibitem[{\citenamefont{Yang et~al.}(2020)\citenamefont{Yang, You, Wang, Huang,
  Xi, Xu, Cao, Tian, Xu, Dai et~al.}}]{PhysRevMaterials.4.024202}
\bibinfo{author}{\bibfnamefont{H.}~\bibnamefont{Yang}},
  \bibinfo{author}{\bibfnamefont{W.}~\bibnamefont{You}},
  \bibinfo{author}{\bibfnamefont{J.}~\bibnamefont{Wang}},
  \bibinfo{author}{\bibfnamefont{J.}~\bibnamefont{Huang}},
  \bibinfo{author}{\bibfnamefont{C.}~\bibnamefont{Xi}},
  \bibinfo{author}{\bibfnamefont{X.}~\bibnamefont{Xu}},
  \bibinfo{author}{\bibfnamefont{C.}~\bibnamefont{Cao}},
  \bibinfo{author}{\bibfnamefont{M.}~\bibnamefont{Tian}},
  \bibinfo{author}{\bibfnamefont{Z.-A.} \bibnamefont{Xu}},
  \bibinfo{author}{\bibfnamefont{J.}~\bibnamefont{Dai}}, \bibnamefont{et~al.},
  \bibinfo{journal}{Phys. Rev. Mater.} \textbf{\bibinfo{volume}{4}},
  \bibinfo{pages}{024202} (\bibinfo{year}{2020}).

\bibitem[{\citenamefont{Ding et~al.}(2019)\citenamefont{Ding, Koo, Xu, Li, Lu,
  Zhao, Wang, Yin, Lei, Yan et~al.}}]{PhysRevX.9.041061}
\bibinfo{author}{\bibfnamefont{L.}~\bibnamefont{Ding}},
  \bibinfo{author}{\bibfnamefont{J.}~\bibnamefont{Koo}},
  \bibinfo{author}{\bibfnamefont{L.}~\bibnamefont{Xu}},
  \bibinfo{author}{\bibfnamefont{X.}~\bibnamefont{Li}},
  \bibinfo{author}{\bibfnamefont{X.}~\bibnamefont{Lu}},
  \bibinfo{author}{\bibfnamefont{L.}~\bibnamefont{Zhao}},
  \bibinfo{author}{\bibfnamefont{Q.}~\bibnamefont{Wang}},
  \bibinfo{author}{\bibfnamefont{Q.}~\bibnamefont{Yin}},
  \bibinfo{author}{\bibfnamefont{H.}~\bibnamefont{Lei}},
  \bibinfo{author}{\bibfnamefont{B.}~\bibnamefont{Yan}}, \bibnamefont{et~al.},
  \bibinfo{journal}{Phys. Rev. X} \textbf{\bibinfo{volume}{9}},
  \bibinfo{pages}{041061} (\bibinfo{year}{2019}).

\bibitem[{\citenamefont{Geishendorf et~al.}(2020)\citenamefont{Geishendorf,
  Vir, Shekhar, Felser, Facio, van~den Brink, Nielsch, Thomas, and
  Goennenwein}}]{geishendorf2019signatures}
\bibinfo{author}{\bibfnamefont{K.}~\bibnamefont{Geishendorf}},
  \bibinfo{author}{\bibfnamefont{P.}~\bibnamefont{Vir}},
  \bibinfo{author}{\bibfnamefont{C.}~\bibnamefont{Shekhar}},
  \bibinfo{author}{\bibfnamefont{C.}~\bibnamefont{Felser}},
  \bibinfo{author}{\bibfnamefont{J.~I.} \bibnamefont{Facio}},
  \bibinfo{author}{\bibfnamefont{J.}~\bibnamefont{van~den Brink}},
  \bibinfo{author}{\bibfnamefont{K.}~\bibnamefont{Nielsch}},
  \bibinfo{author}{\bibfnamefont{A.}~\bibnamefont{Thomas}}, \bibnamefont{and}
  \bibinfo{author}{\bibfnamefont{S.~T.} \bibnamefont{Goennenwein}},
  \bibinfo{journal}{Nano Lett.} \textbf{\bibinfo{volume}{20}},
  \bibinfo{pages}{300} (\bibinfo{year}{2020}).

\bibitem[{\citenamefont{Nakatsuji et~al.}(2015)\citenamefont{Nakatsuji,
  Kiyohara, and Higo}}]{nakatsuji2015large}
\bibinfo{author}{\bibfnamefont{S.}~\bibnamefont{Nakatsuji}},
  \bibinfo{author}{\bibfnamefont{N.}~\bibnamefont{Kiyohara}}, \bibnamefont{and}
  \bibinfo{author}{\bibfnamefont{T.}~\bibnamefont{Higo}},
  \bibinfo{journal}{Nature} \textbf{\bibinfo{volume}{527}},
  \bibinfo{pages}{212} (\bibinfo{year}{2015}).

\bibitem[{\citenamefont{Ye et~al.}(2018)\citenamefont{Ye, Kang, Liu, Von~Cube,
  Wicker, Suzuki, Jozwiak, Bostwick, Rotenberg, Bell et~al.}}]{ye2018massive}
\bibinfo{author}{\bibfnamefont{L.}~\bibnamefont{Ye}},
  \bibinfo{author}{\bibfnamefont{M.}~\bibnamefont{Kang}},
  \bibinfo{author}{\bibfnamefont{J.}~\bibnamefont{Liu}},
  \bibinfo{author}{\bibfnamefont{F.}~\bibnamefont{Von~Cube}},
  \bibinfo{author}{\bibfnamefont{C.~R.} \bibnamefont{Wicker}},
  \bibinfo{author}{\bibfnamefont{T.}~\bibnamefont{Suzuki}},
  \bibinfo{author}{\bibfnamefont{C.}~\bibnamefont{Jozwiak}},
  \bibinfo{author}{\bibfnamefont{A.}~\bibnamefont{Bostwick}},
  \bibinfo{author}{\bibfnamefont{E.}~\bibnamefont{Rotenberg}},
  \bibinfo{author}{\bibfnamefont{D.~C.} \bibnamefont{Bell}},
  \bibnamefont{et~al.}, \bibinfo{journal}{Nature}
  \textbf{\bibinfo{volume}{555}}, \bibinfo{pages}{638} (\bibinfo{year}{2018}).

\bibitem[{\citenamefont{Sales et~al.}(2014)\citenamefont{Sales, Saparov,
  McGuire, Singh, and Parker}}]{Fe3Sn_sales2014ferromagnetism}
\bibinfo{author}{\bibfnamefont{B.~C.} \bibnamefont{Sales}},
  \bibinfo{author}{\bibfnamefont{B.}~\bibnamefont{Saparov}},
  \bibinfo{author}{\bibfnamefont{M.~A.} \bibnamefont{McGuire}},
  \bibinfo{author}{\bibfnamefont{D.~J.} \bibnamefont{Singh}}, \bibnamefont{and}
  \bibinfo{author}{\bibfnamefont{D.~S.} \bibnamefont{Parker}},
  \bibinfo{journal}{Sci. Rep.} \textbf{\bibinfo{volume}{4}}, \bibinfo{pages}{1}
  (\bibinfo{year}{2014}).

\bibitem[{\citenamefont{Fayyazi et~al.}(2019)\citenamefont{Fayyazi, Skokov,
  Faske, Opahle, Duerrschnabel, Helbig, Soldatov, Rohrmann, Molina-Luna,
  G{\"u}th et~al.}}]{Fe3Sn_fayyazi2019experimental}
\bibinfo{author}{\bibfnamefont{B.}~\bibnamefont{Fayyazi}},
  \bibinfo{author}{\bibfnamefont{K.~P.} \bibnamefont{Skokov}},
  \bibinfo{author}{\bibfnamefont{T.}~\bibnamefont{Faske}},
  \bibinfo{author}{\bibfnamefont{I.}~\bibnamefont{Opahle}},
  \bibinfo{author}{\bibfnamefont{M.}~\bibnamefont{Duerrschnabel}},
  \bibinfo{author}{\bibfnamefont{T.}~\bibnamefont{Helbig}},
  \bibinfo{author}{\bibfnamefont{I.}~\bibnamefont{Soldatov}},
  \bibinfo{author}{\bibfnamefont{U.}~\bibnamefont{Rohrmann}},
  \bibinfo{author}{\bibfnamefont{L.}~\bibnamefont{Molina-Luna}},
  \bibinfo{author}{\bibfnamefont{K.}~\bibnamefont{G{\"u}th}},
  \bibnamefont{et~al.}, \bibinfo{journal}{Acta Mater.}
  \textbf{\bibinfo{volume}{180}}, \bibinfo{pages}{126} (\bibinfo{year}{2019}).

\bibitem[{\citenamefont{Chen et~al.}(2022)\citenamefont{Chen, Minami, Sakai,
  Wang, Feng, Nomoto, Hirayama, Ishii, Koretsune, Arita
  et~al.}}]{chen2022large}
\bibinfo{author}{\bibfnamefont{T.}~\bibnamefont{Chen}},
  \bibinfo{author}{\bibfnamefont{S.}~\bibnamefont{Minami}},
  \bibinfo{author}{\bibfnamefont{A.}~\bibnamefont{Sakai}},
  \bibinfo{author}{\bibfnamefont{Y.}~\bibnamefont{Wang}},
  \bibinfo{author}{\bibfnamefont{Z.}~\bibnamefont{Feng}},
  \bibinfo{author}{\bibfnamefont{T.}~\bibnamefont{Nomoto}},
  \bibinfo{author}{\bibfnamefont{M.}~\bibnamefont{Hirayama}},
  \bibinfo{author}{\bibfnamefont{R.}~\bibnamefont{Ishii}},
  \bibinfo{author}{\bibfnamefont{T.}~\bibnamefont{Koretsune}},
  \bibinfo{author}{\bibfnamefont{R.}~\bibnamefont{Arita}},
  \bibnamefont{et~al.}, \bibinfo{journal}{Sci. Adv.}
  \textbf{\bibinfo{volume}{8}}, \bibinfo{pages}{eabk1480}
  (\bibinfo{year}{2022}).

\bibitem[{\citenamefont{Jain et~al.}(2013)\citenamefont{Jain, Ong, Hautier,
  Chen, Richards, Dacek, Cholia, Gunter, Skinner, Ceder
  et~al.}}]{jain2013commentary}
\bibinfo{author}{\bibfnamefont{A.}~\bibnamefont{Jain}},
  \bibinfo{author}{\bibfnamefont{S.~P.} \bibnamefont{Ong}},
  \bibinfo{author}{\bibfnamefont{G.}~\bibnamefont{Hautier}},
  \bibinfo{author}{\bibfnamefont{W.}~\bibnamefont{Chen}},
  \bibinfo{author}{\bibfnamefont{W.~D.} \bibnamefont{Richards}},
  \bibinfo{author}{\bibfnamefont{S.}~\bibnamefont{Dacek}},
  \bibinfo{author}{\bibfnamefont{S.}~\bibnamefont{Cholia}},
  \bibinfo{author}{\bibfnamefont{D.}~\bibnamefont{Gunter}},
  \bibinfo{author}{\bibfnamefont{D.}~\bibnamefont{Skinner}},
  \bibinfo{author}{\bibfnamefont{G.}~\bibnamefont{Ceder}},
  \bibnamefont{et~al.}, \bibinfo{journal}{APL Mater.}
  \textbf{\bibinfo{volume}{1}}, \bibinfo{pages}{011002} (\bibinfo{year}{2013}).

\bibitem[{\citenamefont{Buschow et~al.}(1983)\citenamefont{Buschow, van Engen,
  and Jongebreur}}]{buschow1983magneto}
\bibinfo{author}{\bibfnamefont{K.}~\bibnamefont{Buschow}},
  \bibinfo{author}{\bibfnamefont{P.}~\bibnamefont{van Engen}},
  \bibnamefont{and}
  \bibinfo{author}{\bibfnamefont{R.}~\bibnamefont{Jongebreur}},
  \bibinfo{journal}{J. Magn. Magn. Mater.} \textbf{\bibinfo{volume}{38}},
  \bibinfo{pages}{1} (\bibinfo{year}{1983}).

\bibitem[{\citenamefont{Giefers and Nicol}(2006)}]{FeSn_phasediagram}
\bibinfo{author}{\bibfnamefont{H.}~\bibnamefont{Giefers}} \bibnamefont{and}
  \bibinfo{author}{\bibfnamefont{M.}~\bibnamefont{Nicol}}, \bibinfo{journal}{J.
  Alloy. Compd.} \textbf{\bibinfo{volume}{422}}, \bibinfo{pages}{132}
  (\bibinfo{year}{2006}).

\bibitem[{SM()}]{SM}
\bibinfo{note}{See Supplemental Material at [href] for detailed information
  about the powder X-ray diffraction, Weyl nodes distribution in the Brillouin
  zone, and Berry curvature.}

\bibitem[{\citenamefont{Koepernik and Eschrig}(1999)}]{koepernik1999full}
\bibinfo{author}{\bibfnamefont{K.}~\bibnamefont{Koepernik}} \bibnamefont{and}
  \bibinfo{author}{\bibfnamefont{H.}~\bibnamefont{Eschrig}},
  \bibinfo{journal}{Phys. Rev. B} \textbf{\bibinfo{volume}{59}},
  \bibinfo{pages}{1743} (\bibinfo{year}{1999}).

\bibitem[{fpl()}]{fplo}
\emph{\bibinfo{title}{Fplo}}, \urlprefix\url{https://www.fplo.de/}.

\bibitem[{\citenamefont{Perdew et~al.}(1996)\citenamefont{Perdew, Burke, and
  Ernzerhof}}]{perdew1996generalized}
\bibinfo{author}{\bibfnamefont{J.~P.} \bibnamefont{Perdew}},
  \bibinfo{author}{\bibfnamefont{K.}~\bibnamefont{Burke}}, \bibnamefont{and}
  \bibinfo{author}{\bibfnamefont{M.}~\bibnamefont{Ernzerhof}},
  \bibinfo{journal}{Phys. Rev. Lett.} \textbf{\bibinfo{volume}{77}},
  \bibinfo{pages}{3865} (\bibinfo{year}{1996}).

\bibitem[{\citenamefont{Lejaeghere et~al.}(2016)\citenamefont{Lejaeghere,
  Bihlmayer, Bj{\"o}rkman, Blaha, Bl{\"u}gel, Blum, Caliste, Castelli, Clark,
  Dal~Corso et~al.}}]{lejaeghere2016reproducibility}
\bibinfo{author}{\bibfnamefont{K.}~\bibnamefont{Lejaeghere}},
  \bibinfo{author}{\bibfnamefont{G.}~\bibnamefont{Bihlmayer}},
  \bibinfo{author}{\bibfnamefont{T.}~\bibnamefont{Bj{\"o}rkman}},
  \bibinfo{author}{\bibfnamefont{P.}~\bibnamefont{Blaha}},
  \bibinfo{author}{\bibfnamefont{S.}~\bibnamefont{Bl{\"u}gel}},
  \bibinfo{author}{\bibfnamefont{V.}~\bibnamefont{Blum}},
  \bibinfo{author}{\bibfnamefont{D.}~\bibnamefont{Caliste}},
  \bibinfo{author}{\bibfnamefont{I.~E.} \bibnamefont{Castelli}},
  \bibinfo{author}{\bibfnamefont{S.~J.} \bibnamefont{Clark}},
  \bibinfo{author}{\bibfnamefont{A.}~\bibnamefont{Dal~Corso}},
  \bibnamefont{et~al.}, \bibinfo{journal}{Science}
  \textbf{\bibinfo{volume}{351}}, \bibinfo{pages}{aad3000}
  (\bibinfo{year}{2016}).

\bibitem[{\citenamefont{Koepernik et~al.}(2016)\citenamefont{Koepernik,
  Kasinathan, Efremov, Khim, Borisenko, B\"uchner, and van~den
  Brink}}]{PhysRevB.93.201101}
\bibinfo{author}{\bibfnamefont{K.}~\bibnamefont{Koepernik}},
  \bibinfo{author}{\bibfnamefont{D.}~\bibnamefont{Kasinathan}},
  \bibinfo{author}{\bibfnamefont{D.~V.} \bibnamefont{Efremov}},
  \bibinfo{author}{\bibfnamefont{S.}~\bibnamefont{Khim}},
  \bibinfo{author}{\bibfnamefont{S.}~\bibnamefont{Borisenko}},
  \bibinfo{author}{\bibfnamefont{B.}~\bibnamefont{B\"uchner}},
  \bibnamefont{and} \bibinfo{author}{\bibfnamefont{J.}~\bibnamefont{van~den
  Brink}}, \bibinfo{journal}{Phys. Rev. B} \textbf{\bibinfo{volume}{93}},
  \bibinfo{pages}{201101} (\bibinfo{year}{2016}).

\bibitem[{\citenamefont{Wang et~al.}(2006)\citenamefont{Wang, Yates, Souza, and
  Vanderbilt}}]{wang2006ab}
\bibinfo{author}{\bibfnamefont{X.}~\bibnamefont{Wang}},
  \bibinfo{author}{\bibfnamefont{J.~R.} \bibnamefont{Yates}},
  \bibinfo{author}{\bibfnamefont{I.}~\bibnamefont{Souza}}, \bibnamefont{and}
  \bibinfo{author}{\bibfnamefont{D.}~\bibnamefont{Vanderbilt}},
  \bibinfo{journal}{Phys. Rev. B} \textbf{\bibinfo{volume}{74}},
  \bibinfo{pages}{195118} (\bibinfo{year}{2006}).

\bibitem[{\citenamefont{Onoda et~al.}(2008)\citenamefont{Onoda, Sugimoto, and
  Nagaosa}}]{onoda2008quantum}
\bibinfo{author}{\bibfnamefont{S.}~\bibnamefont{Onoda}},
  \bibinfo{author}{\bibfnamefont{N.}~\bibnamefont{Sugimoto}}, \bibnamefont{and}
  \bibinfo{author}{\bibfnamefont{N.}~\bibnamefont{Nagaosa}},
  \bibinfo{journal}{Phys. Rev. B} \textbf{\bibinfo{volume}{77}},
  \bibinfo{pages}{165103} (\bibinfo{year}{2008}).

\bibitem[{\citenamefont{Nagaosa et~al.}(2010)\citenamefont{Nagaosa, Sinova,
  Onoda, MacDonald, and Ong}}]{nagaosa2010anomalous}
\bibinfo{author}{\bibfnamefont{N.}~\bibnamefont{Nagaosa}},
  \bibinfo{author}{\bibfnamefont{J.}~\bibnamefont{Sinova}},
  \bibinfo{author}{\bibfnamefont{S.}~\bibnamefont{Onoda}},
  \bibinfo{author}{\bibfnamefont{A.~H.} \bibnamefont{MacDonald}},
  \bibnamefont{and} \bibinfo{author}{\bibfnamefont{N.~P.} \bibnamefont{Ong}},
  \bibinfo{journal}{Rev. Mod. Phys.} \textbf{\bibinfo{volume}{82}},
  \bibinfo{pages}{1539} (\bibinfo{year}{2010}).

\bibitem[{\citenamefont{Vekilova et~al.}(2019)\citenamefont{Vekilova, Fayyazi,
  Skokov, Gutfleisch, Echevarria-Bonet, Barandiar\'an, Kovacs, Fischbacher,
  Schrefl, Eriksson et~al.}}]{PhysRevB.99.024421}
\bibinfo{author}{\bibfnamefont{O.~Y.} \bibnamefont{Vekilova}},
  \bibinfo{author}{\bibfnamefont{B.}~\bibnamefont{Fayyazi}},
  \bibinfo{author}{\bibfnamefont{K.~P.} \bibnamefont{Skokov}},
  \bibinfo{author}{\bibfnamefont{O.}~\bibnamefont{Gutfleisch}},
  \bibinfo{author}{\bibfnamefont{C.}~\bibnamefont{Echevarria-Bonet}},
  \bibinfo{author}{\bibfnamefont{J.~M.} \bibnamefont{Barandiar\'an}},
  \bibinfo{author}{\bibfnamefont{A.}~\bibnamefont{Kovacs}},
  \bibinfo{author}{\bibfnamefont{J.}~\bibnamefont{Fischbacher}},
  \bibinfo{author}{\bibfnamefont{T.}~\bibnamefont{Schrefl}},
  \bibinfo{author}{\bibfnamefont{O.}~\bibnamefont{Eriksson}},
  \bibnamefont{et~al.}, \bibinfo{journal}{Phys. Rev. B}
  \textbf{\bibinfo{volume}{99}}, \bibinfo{pages}{024421}
  (\bibinfo{year}{2019}).

\bibitem[{\citenamefont{Nordstr\"om et~al.}(1992)\citenamefont{Nordstr\"om,
  Brooks, and Johansson}}]{nordstrom1992calculation}
\bibinfo{author}{\bibfnamefont{L.}~\bibnamefont{Nordstr\"om}},
  \bibinfo{author}{\bibfnamefont{M.}~\bibnamefont{Brooks}}, \bibnamefont{and}
  \bibinfo{author}{\bibfnamefont{B.}~\bibnamefont{Johansson}},
  \bibinfo{journal}{J. Phys.: Condens. Matter.} \textbf{\bibinfo{volume}{4}},
  \bibinfo{pages}{3261} (\bibinfo{year}{1992}).

\bibitem[{\citenamefont{Ghimire et~al.}(2019)\citenamefont{Ghimire, Facio, You,
  Ye, Checkelsky, Fang, Kaxiras, Richter, and van~den
  Brink}}]{ghimire2019creating}
\bibinfo{author}{\bibfnamefont{M.~P.} \bibnamefont{Ghimire}},
  \bibinfo{author}{\bibfnamefont{J.~I.} \bibnamefont{Facio}},
  \bibinfo{author}{\bibfnamefont{J.-S.} \bibnamefont{You}},
  \bibinfo{author}{\bibfnamefont{L.}~\bibnamefont{Ye}},
  \bibinfo{author}{\bibfnamefont{J.~G.} \bibnamefont{Checkelsky}},
  \bibinfo{author}{\bibfnamefont{S.}~\bibnamefont{Fang}},
  \bibinfo{author}{\bibfnamefont{E.}~\bibnamefont{Kaxiras}},
  \bibinfo{author}{\bibfnamefont{M.}~\bibnamefont{Richter}}, \bibnamefont{and}
  \bibinfo{author}{\bibfnamefont{J.}~\bibnamefont{van~den Brink}},
  \bibinfo{journal}{Phys. Rev. Res.} \textbf{\bibinfo{volume}{1}},
  \bibinfo{pages}{032044} (\bibinfo{year}{2019}).

\bibitem[{\citenamefont{Ray et~al.}(2022)\citenamefont{Ray, Sadhukhan, Richter,
  Facio, and van~den Brink}}]{ray2022tunable}
\bibinfo{author}{\bibfnamefont{R.}~\bibnamefont{Ray}},
  \bibinfo{author}{\bibfnamefont{B.}~\bibnamefont{Sadhukhan}},
  \bibinfo{author}{\bibfnamefont{M.}~\bibnamefont{Richter}},
  \bibinfo{author}{\bibfnamefont{J.~I.} \bibnamefont{Facio}}, \bibnamefont{and}
  \bibinfo{author}{\bibfnamefont{J.}~\bibnamefont{van~den Brink}},
  \bibinfo{journal}{npj Quantum Mater.} \textbf{\bibinfo{volume}{7}},
  \bibinfo{pages}{1} (\bibinfo{year}{2022}).

\bibitem[{\citenamefont{Li et~al.}(2020)\citenamefont{Li, Zhang, Liang, Ding,
  Chen, Shen, Li, Liu, Xi, Wu et~al.}}]{li2020large}
\bibinfo{author}{\bibfnamefont{H.}~\bibnamefont{Li}},
  \bibinfo{author}{\bibfnamefont{B.}~\bibnamefont{Zhang}},
  \bibinfo{author}{\bibfnamefont{J.}~\bibnamefont{Liang}},
  \bibinfo{author}{\bibfnamefont{B.}~\bibnamefont{Ding}},
  \bibinfo{author}{\bibfnamefont{J.}~\bibnamefont{Chen}},
  \bibinfo{author}{\bibfnamefont{J.}~\bibnamefont{Shen}},
  \bibinfo{author}{\bibfnamefont{Z.}~\bibnamefont{Li}},
  \bibinfo{author}{\bibfnamefont{E.}~\bibnamefont{Liu}},
  \bibinfo{author}{\bibfnamefont{X.}~\bibnamefont{Xi}},
  \bibinfo{author}{\bibfnamefont{G.}~\bibnamefont{Wu}}, \bibnamefont{et~al.},
  \bibinfo{journal}{Phys. Rev. B} \textbf{\bibinfo{volume}{101}},
  \bibinfo{pages}{140409} (\bibinfo{year}{2020}).

\bibitem[{\citenamefont{Ozawa and Nomura}(2022)}]{ozawa2022self}
\bibinfo{author}{\bibfnamefont{A.}~\bibnamefont{Ozawa}} \bibnamefont{and}
  \bibinfo{author}{\bibfnamefont{K.}~\bibnamefont{Nomura}},
  \bibinfo{journal}{Physical Review Materials} \textbf{\bibinfo{volume}{6}},
  \bibinfo{pages}{024202} (\bibinfo{year}{2022}).

\bibitem[{\citenamefont{Guguchia et~al.}(2021)\citenamefont{Guguchia, Zhou,
  Wang, Yin, Mielke~III, Tsirkin, Belopolski, Zhang, Cochran, Neupert
  et~al.}}]{guguchia2021multiple}
\bibinfo{author}{\bibfnamefont{Z.}~\bibnamefont{Guguchia}},
  \bibinfo{author}{\bibfnamefont{H.}~\bibnamefont{Zhou}},
  \bibinfo{author}{\bibfnamefont{C.}~\bibnamefont{Wang}},
  \bibinfo{author}{\bibfnamefont{J.-X.} \bibnamefont{Yin}},
  \bibinfo{author}{\bibfnamefont{C.}~\bibnamefont{Mielke~III}},
  \bibinfo{author}{\bibfnamefont{S.}~\bibnamefont{Tsirkin}},
  \bibinfo{author}{\bibfnamefont{I.}~\bibnamefont{Belopolski}},
  \bibinfo{author}{\bibfnamefont{S.-S.} \bibnamefont{Zhang}},
  \bibinfo{author}{\bibfnamefont{T.}~\bibnamefont{Cochran}},
  \bibinfo{author}{\bibfnamefont{T.}~\bibnamefont{Neupert}},
  \bibnamefont{et~al.}, \bibinfo{journal}{Npj Quantum Materials}
  \textbf{\bibinfo{volume}{6}}, \bibinfo{pages}{50} (\bibinfo{year}{2021}).

\bibitem[{\citenamefont{Shen et~al.}(2022)\citenamefont{Shen, Samathrakis, Hu,
  Singh, Fortunato, Liu, Gutfleisch, and Zhang}}]{shen2022thermodynamical}
\bibinfo{author}{\bibfnamefont{C.}~\bibnamefont{Shen}},
  \bibinfo{author}{\bibfnamefont{I.}~\bibnamefont{Samathrakis}},
  \bibinfo{author}{\bibfnamefont{K.}~\bibnamefont{Hu}},
  \bibinfo{author}{\bibfnamefont{H.~K.} \bibnamefont{Singh}},
  \bibinfo{author}{\bibfnamefont{N.}~\bibnamefont{Fortunato}},
  \bibinfo{author}{\bibfnamefont{H.}~\bibnamefont{Liu}},
  \bibinfo{author}{\bibfnamefont{O.}~\bibnamefont{Gutfleisch}},
  \bibnamefont{and} \bibinfo{author}{\bibfnamefont{H.}~\bibnamefont{Zhang}},
  \bibinfo{journal}{npj Comput. Mater.} \textbf{\bibinfo{volume}{8}},
  \bibinfo{pages}{248} (\bibinfo{year}{2022}).

\bibitem[{\citenamefont{Ghimire et~al.}(2021)\citenamefont{Ghimire, Belbase,
  Karki, Ye, You, Facio, Checkelsky, Richter, and Van~den
  Brink}}]{ghimire2021large}
\bibinfo{author}{\bibfnamefont{M.~P.} \bibnamefont{Ghimire}},
  \bibinfo{author}{\bibfnamefont{B.~P.} \bibnamefont{Belbase}},
  \bibinfo{author}{\bibfnamefont{B.}~\bibnamefont{Karki}},
  \bibinfo{author}{\bibfnamefont{L.}~\bibnamefont{Ye}},
  \bibinfo{author}{\bibfnamefont{J.-S.} \bibnamefont{You}},
  \bibinfo{author}{\bibfnamefont{J.}~\bibnamefont{Facio}},
  \bibinfo{author}{\bibfnamefont{J.}~\bibnamefont{Checkelsky}},
  \bibinfo{author}{\bibfnamefont{M.}~\bibnamefont{Richter}}, \bibnamefont{and}
  \bibinfo{author}{\bibfnamefont{J.}~\bibnamefont{Van~den Brink}},
  \bibinfo{journal}{Bull. Am. Phys. Soc.} \textbf{\bibinfo{volume}{66}},
  \bibinfo{pages}{M51.00007} (\bibinfo{year}{2021}).

\end{thebibliography}
\pagebreak
\onecolumngrid

\clearpage 
\begin{center}
    \Large \textbf{Supporting Information: Large anomalous Hall effect in single crystals of the kagome Weyl ferromagnet Fe$_3$Sn}
\end{center}
\section*{X-ray diffraction}
 This data confirms the phase purity of single crystals of Fe$_3$Sn.
\begin{figure}[ht]
	\includegraphics[width=6.0in]{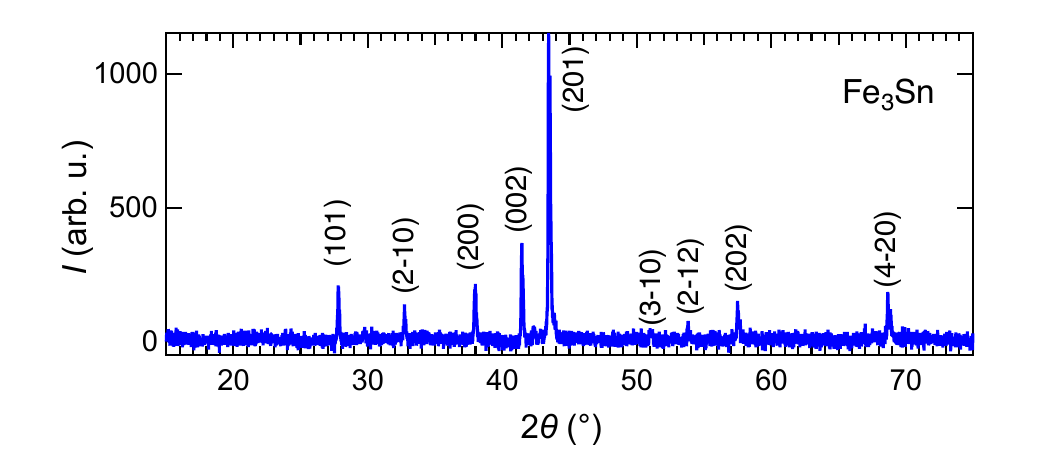}\\
       FIG. S1. Powder XRD pattern of Fe$_3$Sn used in this study (the wavelength of X-ray used is 1.54 \AA).
\end{figure}

\newpage 
\section*{Weyl nodes distribution in Brillouin zone}
In the main text, we have presented the energy of the Weyl nodes, as illustrated in Fig. 4. Additionally, to provide further insight, we depict the distribution of these Weyl nodes in the Brillouin zone in Fig. S2. Weyl nodes closest to Fermi level are found 8 meV below the Fermi energy for the magnetization along [100] direction. There are eight such Weyl nodes connected by symmetry which we present in Fig. S3. 
 
\begin{figure*}[htb!]
\includegraphics[width=5.0in]{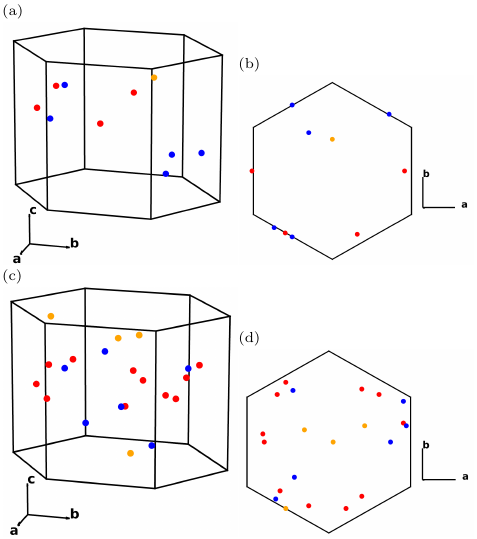}\\
 FIG. S2. Weyl nodes distribution in the three-dimensional (3D) and two-dimensional (2D) Brillouin zone. (a) In 3D for Magnetization along [001] direction (b) In 2D for Magnetization along [001] direction (c) In 3D for Magnetization along [100] direction (d) In 2D for Magnetization along [100] direction. Red, blue and orange dots correspond to Weyl nodes in which the lower band has index $N-1$, $N$ and $N+1$, respectively, with $N$ the number of valence electrons.
\end{figure*}

\begin{figure*}[!h]
	\includegraphics[width=4.0in]{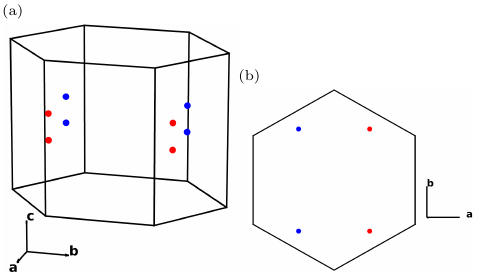}\\
       FIG. S3. Distribution of eight Weyl nodes located at 8 meV below the Fermi level for magnetization along [100] direction  (a) In 3D (b) In 2D. Red and blue dots correspond to Weyl nodes with positive and negative chirality, respectively. 
	 
\end{figure*}

 \newpage

\section*{Anomalous Hall Conductivity (AHC) and Berry curvature}
We have analyzed how states at different parts of the Brillouin zone contribute to the AHC and found that the largest contributions evidence a loop-like structure that reflects weakly gapped nodal lines rather than Weyl nodes as the origin.
   \begin{figure}[ht]
   	\includegraphics[width=6.0in]{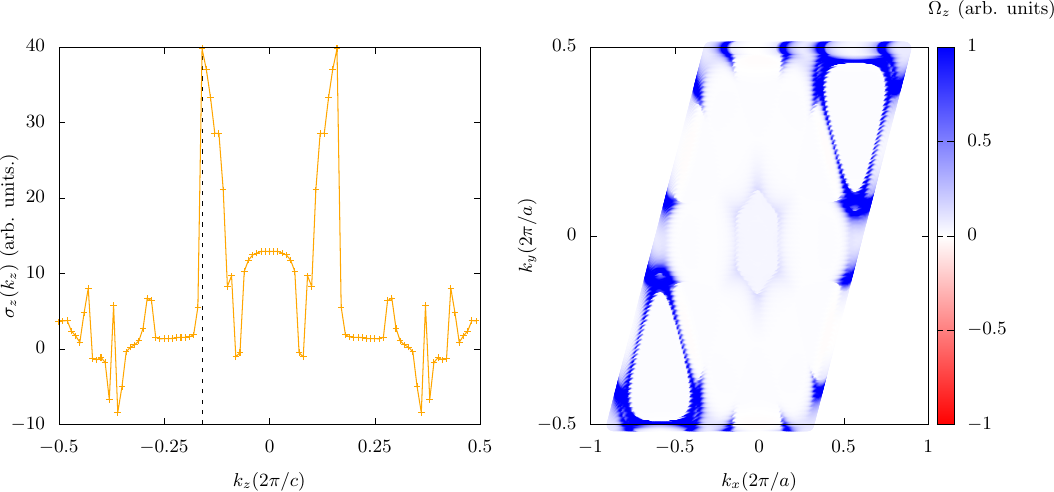}\\
    FIG. S4. Left: Contribution to the AHC from the different planes perpendicular to kz. Right:  $\Omega_z$ as a function of $k_x$ and $k_y$ for the plane that contributes the most to the AHC, indicated by a vertical line in the left panel. The loop-like structure of the hot spots indicates a significant contribution from states other than isolated Weyl nodes. The data corresponds to the Fermi energy and magnetization along [001].
   \end{figure}
\end{document}